\documentclass{revtex4-2}
\usepackage{amsmath}
\usepackage{amssymb}
\usepackage{geometry}
\usepackage{bm}
\usepackage{braket}
\usepackage{color}
\usepackage{graphicx}
\usepackage{todonotes}
\usepackage{aas_macros}
\usepackage{comment}

\definecolor{deepgreen}{rgb}{0.2,0.8,0.2}

\definecolor{gray}{rgb}{0.4,0.4,0.4}

\begin{document}

\date{\today}
\title{ESO White Paper on Intensity Interferometry:\\Cosmology, Fundamental Physics, Quantum Optics}

\author{Robin Kaiser, William Guerin}
\affiliation{Universit\'e C\^ote d’Azur, CNRS, Institut de Physique de Nice, 06200 Nice, France}

\author{Farrokh Vakili}
\affiliation{Universit\'e C\^ote d’Azur, Observatoire de la C\^ote d’Azur, Laboratoire Lagrange, 06000 Nice, France}

\author{Jean-Philippe Berger}
\affiliation{Universit\'e Grenoble Alpes, CNRS, IPAG, 38000 Grenoble, France}

\author{Andrei Nomerotski, Sergei Kulkov, Peter Svihra}
\affiliation{Czech Technical University in Prague, 11519 Prague, Czech Republic}

\author{Eva Santos}
\affiliation{Institute of Physics of the Czech Academy of Sciences, 18200 Prague, Czech Republic}

\author{Colin Carlile, Dainis Dravins}
\affiliation{Lund Observatory, Division of Astrophysics, Department of Physics, Lund University, 22100 Lund, Sweden}

\author{Stefan Funk}
\affiliation{Friedrich-Alexander-Universit\"at Erlangen-N\"urnberg, 91058 Erlangen, Germany}

\author{Prasenjit Saha}
\affiliation{{University of Zurich, Switzerland}}

\author{Roland Walter}
\affiliation{{University of Geneva, Switzerland}}

\author{Marcelo Borges Fernandes}
\affiliation{Observatori\'o Nacional / MCTI, Rio de Janeiro, RJ, Brazil}

\author{Alex G.\ Kim}
\affiliation{Lawrence Berkeley National Laboratory, Berkeley, CA, USA}

\author{David Dunsky, Ken Van Tilburg}
\affiliation{New York University and Flatiron Institute, New York, NY, USA}

\author{Masha Baryakhtar}
\affiliation{University of Washington, Seattle, WA, USA}


\author{Marios Galanis, Robert V.\ Wagoner}
\affiliation{Stanford University, Stanford, CA, USA}

\author{Neal Dalal, Junwu Huang} 
\affiliation{Perimeter Institute for Theoretical Physics, ON, Canada}

\author{Charles Gammie}
\affiliation{University of Illinois Urbana-Champaign}

\author{Norman W.\ Murray} 
\affiliation{Canadian Institute for Theoretical Astrophysics, University of Toronto, ON, Canada}


\maketitle

\clearpage

\setcounter{page}{1}

\section{Introduction}

Intensity interferometry (II) exploits the second-order coherence properties of light from an astrophysical source observed by two or more separated telescopes. Because II is insensitive to atmospheric phase noise and has relatively loose optical-path tolerances, it naturally accommodates very long projected baselines, enabling a robust route to sub-milliarcsecond optical resolution and differential astrometry. Originally invented by Hanbury Brown and Twiss and first demonstrated on Sirius, II culminated in the Narrabri Stellar Intensity Interferometer, which produced a landmark survey of 32 high-precision stellar diameters. 

Modern digital correlators and fast single‑photon detectors developed for quantum technologies now enable a new generation of instruments that revisit and extend this promise to fainter targets. This has led to the revival of II in the modern era, beginning with observations of three bright stars at the Observatoire de la Côte d’Azur in 2018.
A parallel driver of this revival has been the use of Imaging Atmospheric Cherenkov Telescope (IACT) arrays, which provide large light buckets and multi‑hundred‑meter baselines, primarily during bright-Moon periods.
The technique has progressed from proof-of-principle demonstration to routine measurement at different sites, such as VERITAS, MAGIC, and H.E.S.S.\ observatories. The community of II is rapidly expanding, as illustrated by recent workshops~\cite{Ohio,  Porquerolles, Perimeter, Waischenfeld}.

In this whitepaper, we outline how recent technological advances and ongoing developments open qualitatively new science opportunities in cosmology, fundamental physics, and quantum astrophysics. First, II can contribute to one of the most foundational observables in cosmology: the expansion rate of the Universe. Its angular resolution allows it to resolve the angular extent of extragalactic objects such as supernovae (SNe) or quasars; combined with a physical scale local to the source, this yields an angular diameter distance and hence a ``Hubble diagram". Second, the nature of dark matter (DM) can be probed via the astrometric lensing signatures of tiny DM halos. Third, intensity interferometry gives direct access to second-order coherence properties of astrophysical emission, opening a window onto genuinely quantum aspects of astrophysical light.

\section{Scientific Applications}

\subsection{Cosmology}

A persistent tension in measurements of the cosmic expansion rate (the Hubble constant $H_0$) and the quest to constrain the dark‑energy equation of state highlight how difficult it is to improve cosmological inferences once systematic errors dominate. Many leading probes ultimately rest on empirical relations or complex forward models applied to \emph{unresolved} sources: Type~Ia SNe require light‑curve standardization and host/environment corrections; strong‑lens time delays depend on mass‑model assumptions and line‑of‑sight structure. Progress therefore benefits from complementary, calibration‑independent observables. II offers exactly this: direct angular scales at microarcsecond resolution, enabling geometric distance inference that sidesteps flux calibration and many modeling degeneracies.

Spectroscopic binaries provide a parallax-independent route to distance: radial-velocity curves and the orbital period determine the absolute semimajor axis via Kepler’s laws up to an inclination degeneracy, which can be broken with precision astrometry delivered by II.  Comparing the physical and angular scales gives the distance under systematics distinct from trigonometric parallax, and has been demonstrated with amplitude and intensity interferometry. Beyond binaries, II can measure angular sizes of red clump stars and luminous blue variable supergiants, whose physical sizes can be modeled with stellar structure codes.  Modern II can extend such measurements to fainter, more distant systems in our Galaxy (e.g.~the Magellanic Clouds), furnishing independent calibrators for the first rung of the distance ladder.

II can directly measure the apparent angular size $\theta(t)$ of the photosphere and expanding ejecta of individual SNe as a function of time $t$, while spectroscopy constrains the ejecta's homologous expansion at radius $R(t) \propto v_{\rm ej} t$. Their ratio gives a geometric angular‑diameter distance, $D_A = R/\theta$, that can calibrate further rungs of the cosmic distance ladder or construct a purely independent Hubble diagram. With sufficient baseline coverage, determination of 3D morphology and distance is possible for Type Ia and Type IIP SNe.

Active galactic nuclei (AGN) and quasars extend this geometric approach to higher redshifts. Large-baseline II can directly resolve the broad-line emission region (BLR) of AGN at cosmological distances, providing angular morphology scales. Reverberation mapping independently measures the physical extent of the BLR. Combining the two techniques breaks degeneracies inherent in each method, thereby enabling percent-level determination of the angular-diameter distances to these AGN. Because these sources lie well into the Hubble flow (beyond where SN measurements are possible), they offer independent and complementary inferences of $H_0$ with sufficient precision to resolve the Hubble tension.
 
\subsection{Fundamental Physics} \label{sec:fund_phys}

II opens a route to \emph{geometric}, calibration-independent tests of DM microphysics via astrometric weak lensing. Compact subhalos along a line of sight imprint tiny, time‑dependent deflections on background sources—detectable either as accelerations of individual images or as correlated proper‑motion patterns across fields. This program has recently spurred the first astrometric‑lensing constraints on substructure and pipelines built on \textit{Gaia} data releases, which demonstrate the feasibility and the gains expected from more precise catalogs. II contributes a complementary capability --- microarcsecond-class \emph{differential} astrometry on small angular scales --- on a much smaller catalog of bright sources. The same measurements would also extend precise transverse-velocity measurements well beyond the Solar neighborhood, mapping full three-dimensional stellar motions across the Galaxy. Such data would directly probe the Galactic potential and DM halo, constraining clustering, self-interactions, and substructure through the associated density fluctuations and subhalos.

We highlight a concrete new opportunity where II can drive the field into unexplored territory. In strongly lensed quasars, each macro‑image samples a different path through the lens galaxy and intervening structure; small DM halos along these paths induce \emph{stochastic} astrometric fluctuations that are amplified by macro‑lens magnification and whose temporal power spectrum can be predicted. Measuring the evolving separation vectors between image pairs isolates this signal while canceling common systematics. In a forthcoming preprint, a forward model for the associated power spectral density is developed, showing that the most informative observable is the \emph{differential angular acceleration} between image pairs. The lightest subhalos whose crossing time matches the survey duration will dominate this observable. For a bright quadruply-imaged system, e.g.~B1422+231, microarcsecond‑precision differential astrometry over multi‑year observation periods could probe $10^{-6}$--$10^{-2}\,M_\odot$ halos --- masses far below current strong‑lensing and stellar‑stream sensitivities --- while stellar microlensing noise can be subtracted using image‑shape variability and photometry (leaving a bounded residual acceleration floor). These results point to II as a discovery‑space instrument for substellar‑mass structure and, by extension, for the DM kinetic‑decoupling scale and transfer‑function cutoffs shaped by particle microphysics.

\subsection{Quantum Astrophysics}

The introduction of spectroscopy in astrophysics has been transformational, enabling access to a wealth of information not available from photometric observations alone. In addition, neutrino and gravitational-wave detection have opened further observational windows onto the Universe. 

In parallel with these developments in astrophysics, quantum technologies have made spectacular progress over the last decade, from fundamental physics to applied science.
We are therefore at an opportune moment to bring quantum physics --- in particular quantum optics --- into astrophysics, adding yet another messenger to our exploration of the Universe.
Distinctive observational tools in quantum optics are based on correlation measurements, as exemplified by tests of Bell’s inequalities or the experimental characterization of single-photon sources. Moreover, the pioneering work of R.~Glauber showed that second-order coherence, as measured by intensity correlation  functions, emerges as a qualitatively new phenomenon that cannot be inferred from field correlation functions alone, such as those encoded in optical spectra. 

Many laboratory experiments illustrate these concepts, with laser emission based on stimulated emission being the most prominent example. Such systems require gain, typically based on population inversions, and some form of feedback, typically provided by optical cavities. Whereas population inversion and gain are expected to exist in astrophysical environments, feedback must be realized by a different mechanism. One such mechanism is ``random lasing'', where multiple scattering provides sufficient feedback to cross a lasing threshold in the absence of mirrors. It is also possible to modify the second-order coherence function without such feedback, in regimes where amplified spontaneous emission is sufficiently strong to saturate.

Stimulated emission is believed to underlie the emission process of astrophysical masers. A definitive confirmation of the underlying mechanism, however, requires the detection of intensity correlations, which is more readily carried out in the optical regime where efficient and even single-photon detectors are now available. 
One prominent first target for investigating second-order coherence is provided by certain forbidden lines in the Weigelt blobs of Eta Carinae; some Wolf-Rayet, luminous blue variables, and B[e] stars have also been identified as potential sources of non-classical light. The instrumental implementation of second-order coherence measurements closely follows the Hanbury Brown-Twiss technique but requires only a single telescope.
The investigation of stimulated processes will also impact radiative-transfer models, with optical forces no longer restricted to repulsive radiation pressure. The field of laser-cooled atoms has explored such forces in great detail, and a substantial body of expertise exists to adapt these concepts to astrophysical conditions. 
An ideal long-term goal would be a systematic survey facility, similar to 4MOST on VISTA, equipped with a large number of optical fibers feeding intensity-correlation measurements over a wide field of view and in many spectral channels.

\section{Technical Requirements}

Currently, the highest resolution in astronomical imaging is achieved by radio interferometers operating at their highest frequencies.  In the optical band, phase/amplitude interferometers such as VLTI reach baselines of a few hundred meters,  enabling milli-arcsecond-scale structures to be resolved and yielding tantalizing results. The stability requirements for such classical optical amplitude interferometers, however, limit their operation over longer baselines, especially at shorter visual wavelengths, if they are to compete with or surpass radio interferometers such as the Event Horizon Telescope. This limitation has prompted searches for advanced solutions such as free-flying interferometers in space or arrays placed on the Moon. Such proposals are highly challenging for the immediate future, but there is an easier route. As we propose in this whitepaper, II can bypass these limitations and open a window onto important new science cases in cosmology, fundamental physics, and (quantum) astrophysics. 

II is rapidly becoming an active area of research again, galvanized by technological developments that allow for resolving astronomical objects far dimmer than those observed by Hanbury Brown and Twiss. Historically, the main drawback of II has been the requirement of large light-collecting areas and long observation times to reach a high signal-to-noise ratio (SNR). Vast improvements in detector technology, from fast photodetectors to high-resolution spectral multiplexing, combined with large arrays and longer baselines, can significantly increase the SNR and permit smaller, dimmer objects to be resolved. Indeed, these measures are now being applied by several international efforts, such as the Multi-Aperture Spectroscopic Telescope (MAST), QUASAR, the Large Fiber Array Spectroscopic Telescope (LFAST), as well as IACTs under construction, such as the Cherenkov Telescope Array (CTA) and the ASTRI Mini-Array. 

Recent design work shows that modest hardware extensions can expand II’s effective field of view from the diffraction-limited coherence patch to the atmospheric isoplanatic angle, while retaining its angular resolution and light-centroiding performance. Such ``extended-path'' schemes multiplex intensity fluctuations from multiple sky positions onto common detectors, enabling arcsecond-scale differential astrometry with microarcsecond-level precision over the isoplanatic patch. This modification is essential for the differential-astrometry applications in Sec.~\ref{sec:fund_phys} and substantially broadens the science reach of II, from emission morphology measurements to high-precision relative astrometry.

The science applications highlighted in this whitepaper provide strong motivation for continuing II's experimental revival as a worthwhile scientific endeavor.

\bibliography{II_WP}

@misc{Ohio,
  title        = {Stellar Intensity Interferometry 2023, Ohio, USA},
   note         = {\url{https://ccapp.osu.edu/workshops/SII2023} }
}

@misc{Porquerolles,
  title        = {Stellar Intensity Interferometry 2024, Porquerolles, France},
   note         = {\url{https://inphyni.univ-cotedazur.eu/sites/cold-atoms/stellar-intensity-interferometry-2024-workshop} }
}

@misc{Perimeter,
  title        = {Future Prospects of Intensity Interferometry 2024, Perimeter Institute, Canada},
   note         = {\url{https://events.perimeterinstitute.ca/event/347/} }
}

@misc{Waischenfeld,
  title        = {Stellar Intensity Interferometry 2025, Waischenfeld, Germany},
   note         = {\url{https://indico.ecap.work/event/129/} }
}

\end{document}